\documentclass[12pt,dvips]{article}
\usepackage{graphicx}
\usepackage{amssymb}
\usepackage{latexsym}
\usepackage{times}
\usepackage{amsfonts}

\begin{document}
\begin{titlepage}
\setcounter{page}{1}
\renewcommand{\thefootnote}{\fnsymbol{footnote}}

\vspace{5mm}

\begin{center}

{\Large {\bf On mutually unbiased bases: Passing from $d$ to $d^2$ dimensions}}

\baselineskip= 0.80 true cm

\vspace{1.5cm}

{\bf M. R. Kibler$^{1,2,3}$}

\vspace{0.8cm}

$^1$Universit\'e de Lyon, 37 rue du repos, 69361 Lyon, France 

$^2$Universit\'e Claude Bernard Lyon 1, 43 Bd du 11 Novembre 1918, 69622 Villeurbanne, France 

$^3$CNRS/IN2P3, Institut de Physique Nucl\'eaire, 4 rue Enrico Fermi, 69622 Villeurbanne, France 

\vspace{0.5cm} 

E-mail : m.kibler@ipnl.in2p3.fr

\end{center}

\vspace{1.5cm}

\begin{abstract}
We show how to transform the problem of finding $d + 1$ mutually unbiased bases in 
$\mathbb{C}^{d}$ into the one of finding $d(d + 1)$ vectors in $\mathbb{C}^{d^2}$. 
The transformation formulas admit a solution when $d$ is a prime number. 

\bigskip


\end{abstract}

\end{titlepage}

\newpage

\section{Introduction}
The determination of mutually unbiased bases (MUBs) is of pivotal importance in the theory 
of information and in finite quantum mechanics. Let us recall that two orthonormal bases of a 
unitary space are said to be unbiased if the modulus of the inner product of any vector 
of one basis with any vector of the other is independent of the considered vectors (see 
Section II for a definition of MUBs in $\mathbb{C}^{d}$). Such bases are useful in classical 
information (network communication protocols) \cite{1Calderbank} and quantum information 
(quantum state tomography and quantum cryptography) \cite{2Cerf} as well as for the 
construction of discrete Wigner functions \cite{3Gibbons}, the solution of the mean 
King problem \cite{4Englert} and the understanding of the Feynman path integral 
formalism \cite{5Tolar}. They are at the root of a formulation of the Bohr complementarity 
principle for finite quantum systems. 

There exist numerous ways of constructing sets of MUBs. Most of them are based on
discrete Fourier transform over Galois fields and Galois rings, quadratic discrete 
Fourier transform of qudits, discrete Wigner distribution, generalized Pauli operators, 
generalized Hadamard matrices, mutually orthogonal Latin squares, finite geometry methods, 
angular momentum theory, Lie-like approaches, and phase states associated with a generalized 
Weyl-Heisenberg algebra (see \cite{6Durt} and \cite{7Kibler2009} for a review on the subject).

The aim of this note is to introduce a transformation that makes possible to replace the search 
of $d+1$ MUBs in $\mathbb{C}^d$ by the determination of $d(d+1)$ vectors in $\mathbb{C}^{d^2}$.  

\section{Mutually unbiased bases in $\mathbb{C}^{d}$}
Two distinct orthonormal bases 
\begin{equation}
B_a = \{ | a \alpha \rangle : \alpha = 0, 1, \ldots, d-1 \}
\label{def Ba}
\end{equation}
and
\begin{equation}
B_b = \{ | b \beta  \rangle : \beta  = 0, 1, \ldots, d-1 \}
\label{def Bb}
\end{equation}
(with $a \not= b$) of the $d$-dimensional Hilbert space $\mathbb{C}^{d}$
($d \geq 2$) are said to be unbiased if 
\begin{equation}
| \langle a\alpha | b\beta \rangle | = \frac{1}{\sqrt{d}},  
\label{def MUBs}
\end{equation}
where $\langle \ | \ \rangle$ denotes the inner product in $\mathbb{C}^{d}$. It is 
well-known that the maximum number of MUBs in $\mathbb{C}^{d}$ is 
$d + 1$ and that a complete set of $d + 1$ MUBs exists if $d$ is prime or 
the power of a prime number \cite{1Calderbank}, \cite{8Ivanovic}, \cite{9WoottersFields89}. On the other hand, it is not known if it is possible 
to construct a complete set of $d+1$ MUBs in $\mathbb{C}^{d}$ in the case where $d$ is not the 
$n$th power ($n \in \mathbb{N}^*$) of a prime. However, in this case there exists at least 
3 MUBs, a well-known result for $d = 6$. In spite of a great number of numerical 
studies, no more than 3 MUBs were obtained for $d = 6$ \cite{10Grassl2005}, \cite{11Bengtsson2007}, \cite{12BrierleyWeigert2009}, \cite{13Weigert}, \cite{14Weigert}, in agreement with the fact that it is widely believed that only 3 MUBs exist for $d = 6$. 

If we include the $a = b$ case, Eq.~(3) leads to 
\begin{equation}
| \langle a\alpha | b\beta \rangle | = 
\delta_{\alpha ,\beta }\delta_{a,b} + \frac{1}{\sqrt{d}}(1-\delta_{a,b})  
\label{complete1 def MUBs}
\end{equation}
or equivalently 
\begin{equation}
| \langle a\alpha | b\beta \rangle |^2 = 
\delta_{\alpha ,\beta }\delta_{a,b} + \frac{1}{d}(1-\delta_{a,b}).  
\label{complete2 def MUBs}
\end{equation}
We note the presence of a modulus in Eqs.~(\ref{complete1 def MUBs}) and (\ref{complete2 def MUBs}). This 
modulus certainly constitutes a handicap when performing numerical calculations. 

\section{Passing from $\mathbb{C}^{d}$ to $\mathbb{C}^{d^2}$}
The problem of finding a complete set of $d + 1$ MUBs in $\mathbb{C}^{d}$ amounts to find $d(d + 1)$ 
vectors $| a\alpha \rangle$ satisfying Eq.~(\ref{complete2 def MUBs}), where $a = 0, 1, \ldots, d$ and 
$\alpha = 0, 1, \ldots, d-1$ (the indexes of type $a$ refer to the bases and, for fixed $a$, the index 
$\alpha $ refers to one of the $d$ vectors of the basis corresponding to $a$). By following the approach 
developed in \cite{16AlbKib} for positive operator valued measures and MUBs, we can transform this 
problem into a (possibly) simpler one (not involving a square 
modulus like in Eq.~(\ref{complete2 def MUBs})). The idea of the transformation is to introduce 
a projection operator associated with the $| a\alpha \rangle$ vector.  

Let us suppose that it is possible to find $d + 1$ sets $B_a$ (with $a = 0, 1, \ldots, d$) of vectors of 
$\mathbb{C}^{d}$ such that Eq.~(\ref{complete2 def MUBs}) is satisfied. It is thus possible to construct 
$d(d + 1)$ projection operators
 \begin{equation}
\Pi _{a\alpha} = | a\alpha \rangle \langle a\alpha |, \quad a = 0, 1, \ldots, d, \quad \alpha = 0, 1, \ldots, d-1.
\label{Piaalpha}
 \end{equation}
From Eqs.~(\ref{complete2 def MUBs}) and (\ref{Piaalpha}), it is clear that the $\Pi _{a\alpha}$ operators (of rank-1)
satisfy the trace conditions
\begin{equation}
{\rm Tr}\left( \Pi_{ a\alpha } \right) = 1, \quad 
{\rm Tr}\left( \Pi_{ a\alpha } \Pi_{ b\beta } \right) = 
\delta_{\alpha ,\beta }\delta_{a,b} + \frac{1}{d}(1-\delta_{a,b}), 
\label{trace}
\end{equation}
where the traces are taken over $\mathbb{C}^{d}$. Each operator $\Pi_{ a\alpha }$ can be developed on an orthonormal 
basis $\{ E_{pq} : p, q = 0, 1, \ldots, d-1 \}$ of the space of linear operators on $\mathbb{C}^{d}$ (orthonormal with 
respect to the Hilbert-Schmidt inner product). In other words 
\begin{equation}
\Pi_{ a\alpha } = \sum_{p=0}^{d-1} \sum_{q=0}^{d-1} w_{pq}(a\alpha) E_{pq}.  
\label{Pi en E} 
\end{equation}
The $E_{pq}$ operators are generators of the $\mathrm{GL}$($d,\mathbb{C}$) complex Lie group. Their main 
properties are 
\begin{equation}
E_{pq}^{\dagger} = E_{qp}, \ 
E_{pq} E_{rs} = \delta_{q,r} E_{ps}, \ 
{\rm Tr}\left( E_{pq} \right) = \delta_{p,q}, \ 
{\rm Tr}\left( E_{pq}^{\dagger} E_{rs} \right) = \delta_{p,r} \delta_{q,s}, \ 
p,q,r,s \in \mathbb{Z}/d\mathbb{Z}
\label{properties of the Epq} 
\end{equation} 
and they can be represented by the projectors 
\begin{equation}
E_{pq} = | p \rangle \langle q |, \quad p,q \in \mathbb{Z}/d\mathbb{Z}.
\label{rep des Epq} 
\end{equation} 
The $w_{pq}(a\alpha)$ expansion coefficients in Eq.~(\ref{Pi en E}) are complex numbers such that 
\begin{equation}
\overline{w_{pq}(a\alpha)} = w_{qp}(a\alpha), \quad p,q \in \mathbb{Z}/d\mathbb{Z},
\label{hermiticity} 
\end{equation} 
where the bar denotes complex conjugation. 

By combining Eqs.~(\ref{trace}) and (\ref{Pi en E}), we get  
\begin{equation}
\sum_{p=0}^{d-1} \sum_{q=0}^{d-1} \overline{w_{pq}(a\alpha)} w_{pq}(b\beta) = 
\delta_{\alpha ,\beta }\delta_{a,b} + \frac{1}{d}(1-\delta_{a,b}).  
\label{MUBs en w} 
\end{equation} 
The $\Pi_{ a\alpha }$ operators can be considered as vectors 
\begin{equation}
{\bf w}(a\alpha) = \left( w_{00}(a\alpha), w_{01}(a\alpha), \ldots, w_{mm}(a\alpha) \right), \quad m=d-1 
\label{def waalpha} 
\end{equation} 
in the Hilbert space $\mathbb{C}^{d^2}$ of dimension $d^2$ endowed with the usual inner product 
\begin{equation}
{\bf w} (a\alpha) \cdot {\bf w}(b\beta) = \sum_{p=0}^{d-1} \sum_{q=0}^{d-1} \overline{w_{pq}(a\alpha)} w_{pq}(b\beta) 
\label{produit scalaire usuel} 
\end{equation} 
(in Eq.~(\ref{def waalpha}), we use the dictionary order for ordering 
the components of ${\bf w} (a\alpha)$). Equation 
(\ref{MUBs en w}) can then be rewritten as 
\begin{equation}
{\bf w} (a\alpha) \cdot {\bf w}(b\beta) = 
\delta_{\alpha ,\beta}\delta_{a,b} + \frac{1}{d}(1-\delta_{a,b}), 
\label{MUBs en wbis} 
\end{equation} 
to be compared with Eq.~(\ref{complete2 def MUBs}). 

The determination of the $\Pi_{ a\alpha }$ operators and, therefore, of the  $| a\alpha \rangle$ vectors in 
$\mathbb{C}^d$, is equivalent to the determination of the $w_{pq}(a\alpha)$ components of the ${\bf w} (a\alpha)$ 
vectors in $\mathbb{C}^{d^2}$. This yields the following. 

\bigskip 

{\bf Proposition 1.} {\it For $d \geq 2$, to find $d + 1$ MUBs in $\mathbb{C}^d$ (if they exist) is equivalent to find $d(d + 1)$ vectors 
${\bf w} (a\alpha)$ in $\mathbb{C}^{d^2}$, of components $w_{pq}(a\alpha)$ such that $w_{pq}(a\alpha) = \overline{w_{qp}(a\alpha)}$ 
(with $p,q = 0, 1, \ldots, d-1$) and $\sum_{p = 0}^{d-1}w_{pp}(a\alpha) = 1$, satisfying  
\begin{equation}
{\bf w} (a\alpha) \cdot {\bf w}(a\beta) = \delta_{\alpha ,\beta}
\label{proposition1a} 
\end{equation}  
and 
\begin{equation}
{\bf w} (a\alpha) \cdot {\bf w}(b\beta) = \frac{1}{d} \ {\rm for} \ a \not= b, 
\label{proposition1b} 
\end{equation} 
where $a, b = 0, 1, \ldots, d$ and $\alpha = 0, 1, \ldots, d-1$.} 

\bigskip 

{\it Proof.} The proof follows from Eqs.~(\ref{Piaalpha})--(\ref{MUBs en wbis}). \hfill $\Box$

\bigskip 

For $a \not= b$, Eqs.~(\ref{proposition1a}) and (\ref{proposition1b}) show that angle $\omega_{a \alpha b \beta}$ 
between any vector ${\bf w}(a\alpha)$ and any vector ${\bf w}(b\beta)$ is 
\begin{equation}
\omega_{a \alpha b \beta} = \cos^{-1}(1/d)
\label{design} 
\end{equation} 
and therefore does not depend on $a$, $\alpha$, $b$ and $\beta$.

Proposition 1 can be transcribed in terms of matrices. Let $M_{a \alpha}$ be the positive-semidefinite matrix 
of dimension $d$ whose elements are ${w_{pq}(a\alpha)}$ with $p,q \in \mathbb{Z}/d\mathbb{Z}$. Then, 
Eq.~(\ref{produit scalaire usuel}) gives
\begin{equation}
{\bf w} (a\alpha) \cdot {\bf w}(b\beta) = {\rm Tr}\left( M_{a \alpha} M_{b \beta} \right).
\label{produit scalaire en M} 
\end{equation}  
Therefore, we have the following proposition.  

\bigskip 

{\bf Proposition 2.} {\it For $d \geq 2$, to find $d + 1$ MUBs in $\mathbb{C}^d$ (if they exist) is equivalent to find $d(d + 1)$ 
positive-semidefinite (and thus Hermitian) matrices $M_{a\alpha}$ of dimension $d$ satisfying 
\begin{equation}
{\rm Tr}\left( M_{a \alpha} \right) = 1, \quad  
{\rm Tr}\left( M_{a \alpha} M_{b \beta} \right) = \delta_{\alpha ,\beta}\delta_{a,b} + \frac{1}{d}(1-\delta_{a,b})
\label{proposition2} 
\end{equation}  
where $a, b = 0, 1, \ldots, d$ and $\alpha, \beta = 0, 1, \ldots, d-1$.} 

\bigskip 

{\it Proof.} The proof is trivial. \hfill $\Box$

\bigskip 

It is to be noted that Proposition 2 is in agreement with the result of Ref.~\cite{before17} according to which a 
complete set of $d+1$ MUBs forms a convex polytope in the set of Hermitian matrices of dimension $d$ and unit trace. 

Finally, as a test of the validity of Propositions 1 and 2, we have the following result.

\bigskip

{\bf Proposition 3.} {\it For $d$ prime, Eqs.~(\ref{proposition1a}) 
and (\ref{proposition1b}) or Eq.~(\ref{proposition2}) admit the solution
\begin{equation}
{w_{pq}(a\alpha)} = \frac{1}{d} {\rm e}^{{\rm i} \pi (p-q)[(d - 2 - p - q)a - 2 \alpha]/d}, \quad 
a, \alpha, p, q \in \mathbb{Z}/d\mathbb{Z}
\label{d valeurs de a} 
\end{equation}  
and 
\begin{equation}
{w_{pq}(d\alpha)} = \delta_{p,q} \delta_{p,\alpha}, \quad 
\alpha, p, q \in \mathbb{Z}/d\mathbb{Z}
\label{a equals d} 
\end{equation}
for $a = d$.} 

\bigskip

{\it Proof.} The proof is based on the use of Gauss sums \cite{17Berndt} in connection with ordinary 
\cite{18Vourdas04} and quadratic \cite{19InTech} discrete Fourier transforms. Indeed, it is sufficient 
to calculate ${\bf w} (a\alpha) \cdot {\bf w}(b\beta)$ as given by (\ref{produit scalaire usuel}) with 
the help of (\ref{d valeurs de a}) and (\ref{a equals d}) in the cases $a = b$ (for $a = 0, 1, \ldots, d$), 
$a \not= b$ (for $a, b = 0, 1, \ldots, d-1$) and $a \not= b$ (for $a = 0, 1, \ldots, d-1$ and $b=d$). The 
main steps are the following.

(i) Case $a = b = d$: We have
\begin{equation}
{\bf w} (d\alpha) \cdot {\bf w}(d\beta) = 
\sum_{p=0}^{d-1} \sum_{q=0}^{d-1} \delta_{p,q} \delta_{p,\alpha} \delta_{p,\beta} = \delta_{\alpha,\beta}. 
\label{cas1} 
\end{equation}

(ii) Case $a = b = 0, 1, \ldots, d-1$: We have
\begin{equation}
{\bf w} (a\alpha) \cdot {\bf w}(a\beta) = \frac{1}{d^2} 
\sum_{p=0}^{d-1} \sum_{q=0}^{d-1} {\rm e}^{{\rm i} 2 \pi (p-q)(\alpha - \beta)/d} = \delta_{\alpha,\beta}. 
\label{cas2} 
\end{equation}

(iii) Case $a \not= b$ ($a = 0, 1, \ldots, d-1$ and $b=d$): We have
\begin{equation}
{\bf w} (a\alpha) \cdot {\bf w}(d\beta) = \frac{1}{d} 
\sum_{p=0}^{d-1} \sum_{q=0}^{d-1} {\rm e}^{-{\rm i} \pi (p-q)[(d - 2 - p - q)a - 2 \alpha]/d} \delta_{p,q} \delta_{p,\alpha} = 
\frac{1}{d} \sum_{p=0}^{d-1} \delta_{p,\alpha} = \frac{1}{d}. 
\label{cas3} 
\end{equation}

(iv) Case $a \not= b$ ($a, b = 0, 1, \ldots, d-1$): We have
\begin{equation}
{\bf w} (a\alpha) \cdot {\bf w}(b\beta) = \frac{1}{d^2} 
\sum_{p=0}^{d-1} \sum_{q=0}^{d-1} {\rm e}^{{\rm i} \pi (p-q)[(d - 2 - p - q)(b-a) + 2 (\alpha-\beta)]/d}. 
\label{cas4} 
\end{equation}
The double sum in (\ref{cas4}) can be factored into the product of two sums. This leads to 
\begin{equation}
{\bf w} (a\alpha) \cdot {\bf w}(b\beta) = \frac{1}{d^2} 
\left\vert \sum_{k=0}^{d-1} {\rm e}^{{\rm i} \pi \{ (a-b) k^2 + [(d-2) (b-a) + 2 (\alpha-\beta)] k \} /d} \right\vert^2. 
\label{une seule somme} 
\end{equation}
By introducing the generalized Gauss sums \cite{17Berndt} 
\begin{equation}
S(u, v, w) = \sum_{k=0}^{\vert w \vert - 1} {\rm e}^{{\rm i} \pi ( u k^2 + v k ) / w},  
\label{Gauss sum} 
\end{equation}
(where $u$, $v$ and $w$ are integers such that $u$ and $w$ are coprime, $uw$ is nonvanishing and $uw + v$ is even), we obtain 
\begin{equation}
{\bf w} (a\alpha) \cdot {\bf w}(b\beta) = \frac{1}{d^2} \vert S(u, v, w) \vert^2, 
\label{ps en terme de S} 
\end{equation}
with 
\begin{equation}
u = a-b, \quad v = - (a-b) (d-2) + 2 (\alpha-\beta), \quad w = d. 
\label{parameters} 
\end{equation}
The $S(u, v, w)$ Gauss sum in (\ref{ps en terme de S})-(\ref{parameters}) can be calculated from the methods in \cite{17Berndt}. This yields 
\begin{equation}
{\bf w} (a\alpha) \cdot {\bf w}(b\beta) = \frac{1}{d}, 
\label{fin cas4} 
\end{equation}
which completes the proof. \hfill $\Box$

\section{Conclusion}
As a conclusion, passing from $\mathbb{C}^d$ to $\mathbb{C}^{d^2}$ amounts to replace the 
square of the modulus of the inner product $\langle a\alpha | b\beta \rangle$ in $\mathbb{C}^d$ 
(see Eq.~(\ref{complete2 def MUBs})) by the inner product ${\bf w} (a\alpha) \cdot {\bf w}(b\beta)$ 
in $\mathbb{C}^{d^2}$ (see Eqs.~(\ref{proposition1a}) and (\ref{proposition1b})). It is expected that 
the determination of the $d(d + 1)$ vectors ${\bf w} (a\alpha)$ satisfying Eqs.~(\ref{proposition1a}) 
and (\ref{proposition1b}) (or the $d(d + 1)$ corresponding matrices $M_{a\alpha}$ satisfying Eq.~(\ref{proposition2})) 
should be easier than the determination of the $d(d + 1)$ vectors $| a\alpha \rangle$ satisfying 
Eq.~(\ref{complete2 def MUBs}). In this respect, the absence of a modulus in (\ref{proposition1b}) 
represents an incremental step. 

Now we may ask the question: How to pass from the ${\bf w} (a\alpha)$ to $\vert a \alpha \rangle$ vectors? Suppose we find 
$d(d+1)$ vectors of type (\ref{def waalpha}) satisfying Eqs.~(\ref{proposition1a}) and (\ref{proposition1b}). Then, the 
$\Pi_{a \alpha}$ operators given by (\ref{Pi en E}) are known. A matrix realization of each $\Pi_{a \alpha}$ operator immediately 
follows from the standard matrix realization of the generators of the $\mathrm{GL}$($d,\mathbb{C}$) group. The eigenvector of 
the matrix of $\Pi_{a \alpha}$ corresponding to the eigenvalue equal to 1 gives the $\vert a \alpha \rangle$ vector. 

Of course, the impossibility of finding  $d(d + 1)$ vectors ${\bf w} (a\alpha)$ 
or $d(d + 1)$ matrices $M_{a\alpha}$ would mean that $d+1$ MUBs do not exist in $\mathbb{C}^d$ when $d$ is 
not a strictly positive power of a prime. 

Transforming a given problem into another one is always interesting even in the case where the new problem does not 
lead to the solution of the first one. In this vein, the existence problem of MUBs in $d$ dimensions was approached 
from the points of view of finite geometry, Latin squares, and Hadamard matrices (see \cite{6Durt} and references 
therein) with some interesting developments. We hope that the results presented here will stimulate further works, 
especially a new way to handle the $d=6$ unsolved problem. 

To close, let us mention that it should be interesting to apply the 
developments in this paper to the concept of {\it weakly} MUBs recently introduced for dealing in the 
$\mathbb{Z}/d\mathbb{Z} \times \mathbb{Z}/d\mathbb{Z}$ phase space \cite{20Shalaby}. 

An extended version of this note will be published elsewhere.

\end{document}